\documentclass[prl,twocolumn,floatfix,nofootinbib,showpacs]{revtex4}

\usepackage[english]{babel}
\usepackage{amsmath,amssymb,bm,slashed}
\usepackage{graphicx}
\usepackage[sort&compress]{natbib}
\usepackage{xcolor}
\usepackage[normalem]{ulem}
\usepackage{hyperref}
\usepackage{subfigure} % for subfigures
\definecolor{red}{rgb}{1.0, 0, 0}

\allowdisplaybreaks

\bibpunct{[}{]}{,}{n}{}{,}

\newcommand{\ev}[1]{\ensuremath{\left\langle #1 %
                     \right\rangle}} % Expectation value
\begin{document}

% =============================================================================
\title{Constraints on dark matter annihilation from AMS-02 results}
\author{Joachim Kopp}  \email[Email: ]{jkopp@mpi-hd.mpg.de}
\affiliation{Max Planck Institut f\"ur Kernphysik, Saupfercheckweg 1, 69117 Heidelberg, Germany \\
}
\date{\today} % FIXME
% =============================================================================

\begin{abstract}
  We use recently released data on the positron-to-electron ratio in cosmic rays
  from the AMS-02 experiment to constrain dark matter annihilation in the Milky Way.
  Due to the yet unexplained positron excess, limits are generally weaker than
  those obtained using other probes, especially gamma rays. This also means
  that explaining the positron excess in terms of dark matter annihilation
  is difficult. Only if very conservative assumptions on the dark matter distribution
  in the Galactic Center region are adopted, it may be possible to accommodate
  dark matter annihilating to leptons with a cross section above $10^{-24}$~cm$^3$/sec.
  We comment on several theoretical mechanisms to explain such large annihilation
  cross sections.
\end{abstract}

%==============================================================================
\begin{flushright}
\end{flushright}

\maketitle
%==============================================================================

Recently, the AMS-02 collaboration has announced a new measurement of the
cosmic ray positron fraction, i.e.\ the flux of positrons, divided by the flux
of electrons plus positrons~\cite{AMS:2013}. AMS-02 confirms an increase in
the positron fraction at energies above $\sim 10$~GeV, which had been observed
previously by PAMELA~\cite{Adriani:2008zr} and
Fermi-LAT~\cite{FermiLAT:2011ab}.  This upturn is difficult to explain by
secondary production of positrons in interactions of other high energy cosmic
rays, but suggests the existence of a yet unidentified galactic source of
positrons~\cite{Serpico:2011wg}.  One interesting candidate are pulsars, fast
rotating neutron stars in whose strong electromagnetic field high energy $e^+e^-$
pairs are produced (see ref.~\cite{Serpico:2011wg} and references therein
for details).

Another possible source of primary positrons is dark matter (DM) annihilation
or decay into high energy Standard Model particles.  This possibility has been
extensively discussed in the context of the PAMELA and Fermi-LAT
results~\cite{Adriani:2008zr, FermiLAT:2011ab}, see for
instance~\cite{Cirelli:2008pk, ArkaniHamed:2008qn, Nardi:2008ix, Bertone:2008xr, Ibe:2008ye,
Meade:2009rb, Mardon:2009rc, Barger:2009yt, Meade:2009iu, Profumo:2009uf,
Cirelli:2009bb, Cline:2010ag, Calvez:2010jq}.  The conclusion is that
explaining the positron excess in terms of dark matter is very difficult: the
required annihilation cross sections are much larger than the ones required to
explain the DM abundance in the Universe through thermal freeze-out.  (Several
mechanisms have been proposed to make the annihilation cross section velocity
dependent and thus different in the early Universe and
today~\cite{ArkaniHamed:2008qn, Ibe:2008ye}.)  Also, one expects the production
of high energy particles to be accompanied by radio signals from synchrotron
radiation and by gamma rays from decays of unstable annihilation products (in
particular $\pi^0 \to \gamma\gamma)$, from final state radiation and from
inverse Compton scattering.  While the magnitude of these signals relative to
the positron signal depends strongly on poorly understood details of galactic
modeling (DM profile, magnetic fields, etc.), particle physics models capable
of explaining the positron excess are often in tension with constraints from
gamma ray and radio observations~\cite{Bertone:2008xr, Cirelli:2009dv, Crocker:2010gy}.

In this note, we discuss dark matter annihilation in view of the new AMS-02
measurements and derive constraints on the DM mass and annihilation cross section.

\paragraph*{Methods.} We take the primary spectrum of positrons from DM annihilation
from~\cite{Cirelli:2010xx, Ciafaloni:2010ti}. For the fluxes after propagation
we use the results from~\cite{Cirelli:2010xx}, as well results we have obtained
with GALPROP~v54~\cite{Moskalenko:1997gh, Strong:1998pw, Ptuskin:2005ax,
Strong:2007nh, Vladimirov:2010aq}, which we have modified to allow for the
inclusion of arbitrary injection spectra.  We have checked that, for identical
propagation models, we can reproduce the results from~\cite{Cirelli:2010xx}
in GALPROP to within $\text{few} \times 10\%$.  We consider the
annihilation channels $\chi\chi \to \mu^+ \mu^-$, $\chi\chi \to \tau^+ \tau^-$,
$\chi\chi \to b\bar{b}$, $\chi\chi \to W^+ W^-$ and $\chi\chi \to ZZ$,
where $\chi$ is the DM
particle.  To illustrate the impact of uncertainties in charged cosmic ray
propagation through the galaxy, we will show limits for the three different
propagation models MIN, MED and MAX from~\cite{Cirelli:2010xx}, originally
introduced in~\cite{Delahaye:2007fr, Donato:2003xg}.  To derive limits on the DM annihilation
cross sections, we compare the theoretically predicted positron flux to the
data from~\cite{AMS:2013}. We use only data points above 20~GeV to be
insensitive to modulation by varying solar activity. We extract the positron flux
$\phi(e^+)$ from the observed positron fraction $\phi(e^+) / [\phi(e^+) +
\phi(e^-)]$ by multiplying the latter with the total electron plus positron
flux $\phi(e^+) + \phi(e^-)$ measured by Fermi-LAT~\cite{Ackermann:2010ij}.
To avoid any reliance on theoretical modelling of the background from secondary
positrons, we set conservative limits by requiring that the predicted positron
flux remains smaller than the measured one (within error bars) at all energies.
More precisely, we define
\begin{align}
  \chi^2 = \sum_i \frac{\big(\phi_{e^+,i}^\text{th} - \phi_{e^+,i}^\text{obs}\big)^2}{\sigma_i^2}
     \theta\big(\phi_{e^+,i}^\text{th} - \phi_{e^+,i}^\text{obs}\big) \,,
  \label{eq:chi2}
\end{align}
where $\phi_{e^+,i}^\text{th}$ and $\phi_{e^+,i}^\text{obs}$ denote the
predicted and observed fluxes, respectively, $\sigma_i$ are the experimental
errors, and the sum runs over energy bins. We set a one-sided 95\% C.L.\ limit
by requiring $\chi^2 < 4.6$.  The relative uncertainties
$\sigma_i/\phi_{e+,i}^\text{obs}$ are conservatively obtained by adding
linearly the uncertainties of the AMS-02 positron fraction and of the Fermi-LAT
$e^+ + e^-$ flux. Within each experimental data set, we have added the quoted
statistical and systematic uncertainties in quadrature.

We will also explore the possibility that the positron excess is
\emph{explained} by DM annihilation.  This requires some assumptions on the
background flux of secondary astrophysical positrons.  Here, we use
GALPROP~v54~\cite{Moskalenko:1997gh, Strong:1998pw, Ptuskin:2005ax,
Strong:2007nh, Vladimirov:2010aq} for both the signal and the background,
with input parameters tuned to reproduce
Fermi's $e^+ + e^-$ spectrum~\cite{Ackermann:2010ij} at energies between
few~GeV and $\text{few} \times 10$~GeV, i.e.\ in an energy range where the $e^+
+ e^-$ flux from annihilation of heavy DM is expected to be negligible. Our
parameter choice is motivated by references~\cite{Ackermann:2010ij,
Strong:Website}. To quantify the uncertainties in the propagation
model, we also show results obtained using GALPROP implementations
of the MIN, MED and MAX models from~\cite{Cirelli:2010xx, Delahaye:2007fr,
Donato:2003xg} discussed above.  To obtain
the preferred parameter regions, we define in analogy to
equation~\eqref{eq:chi2}
\begin{align}
  \chi^2 = \sum_i \frac{\big(\phi_{e^+,i}^\text{th} - \phi_{e^+,i}^\text{obs}\big)^2}{\sigma_i^2} \,, 
  \label{eq:chi2-2}
\end{align}
where $\phi_{e^+,i}^\text{th}$ now includes the signal and background
predictions. We set two-sided ``$3\sigma$''~CL limits by requiring $\chi^2 <
11.8$.  The quotation marks here indicate that the confidence level is based
on statistical errors only because the probability distributions of the systematic
biases in the propagation and backgrounds model are unknown.

\begin{figure*}
  \begin{tabular}{c@{\qquad}c}
    \includegraphics[width=0.40\textwidth]{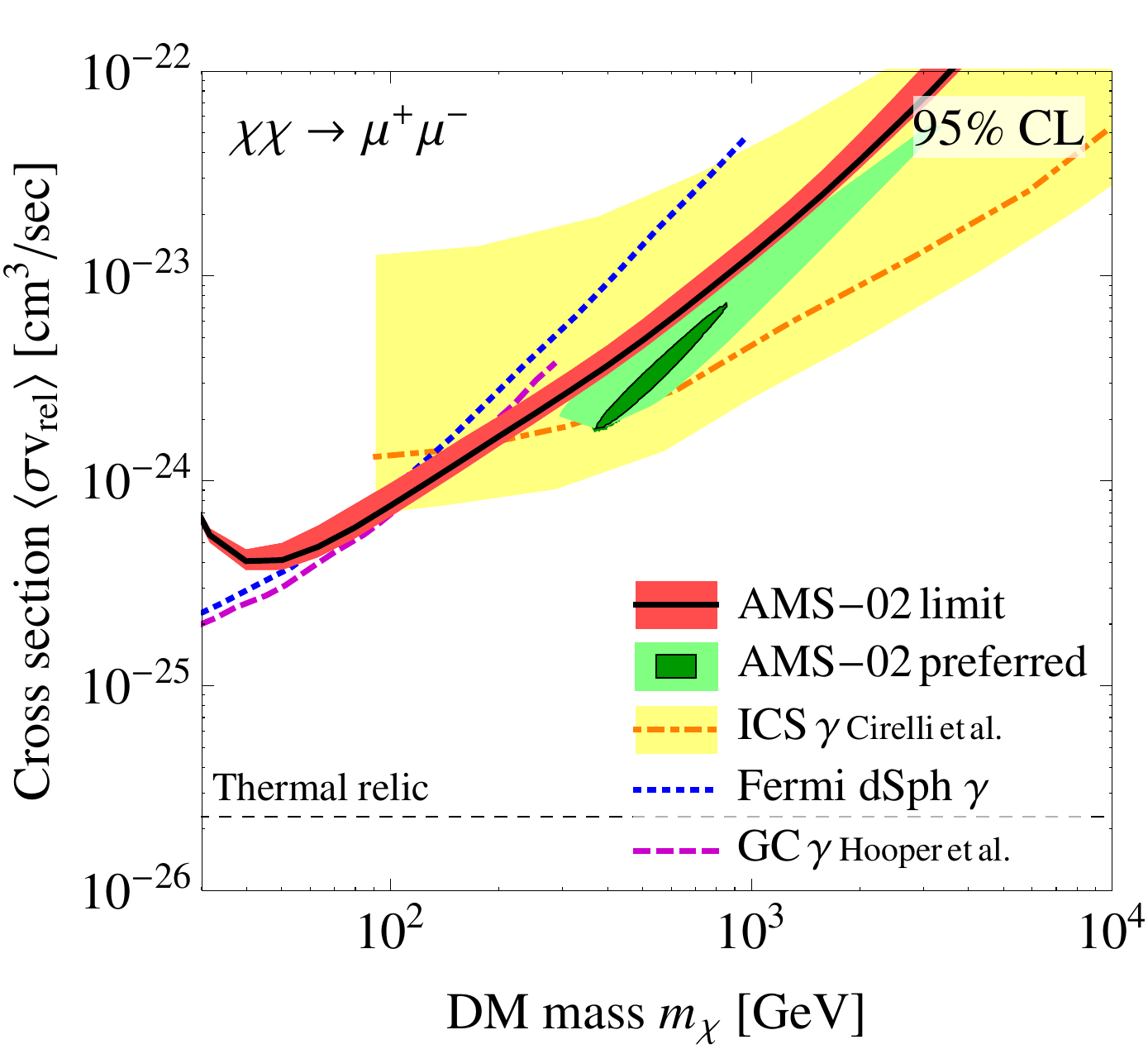} &
    \includegraphics[width=0.40\textwidth]{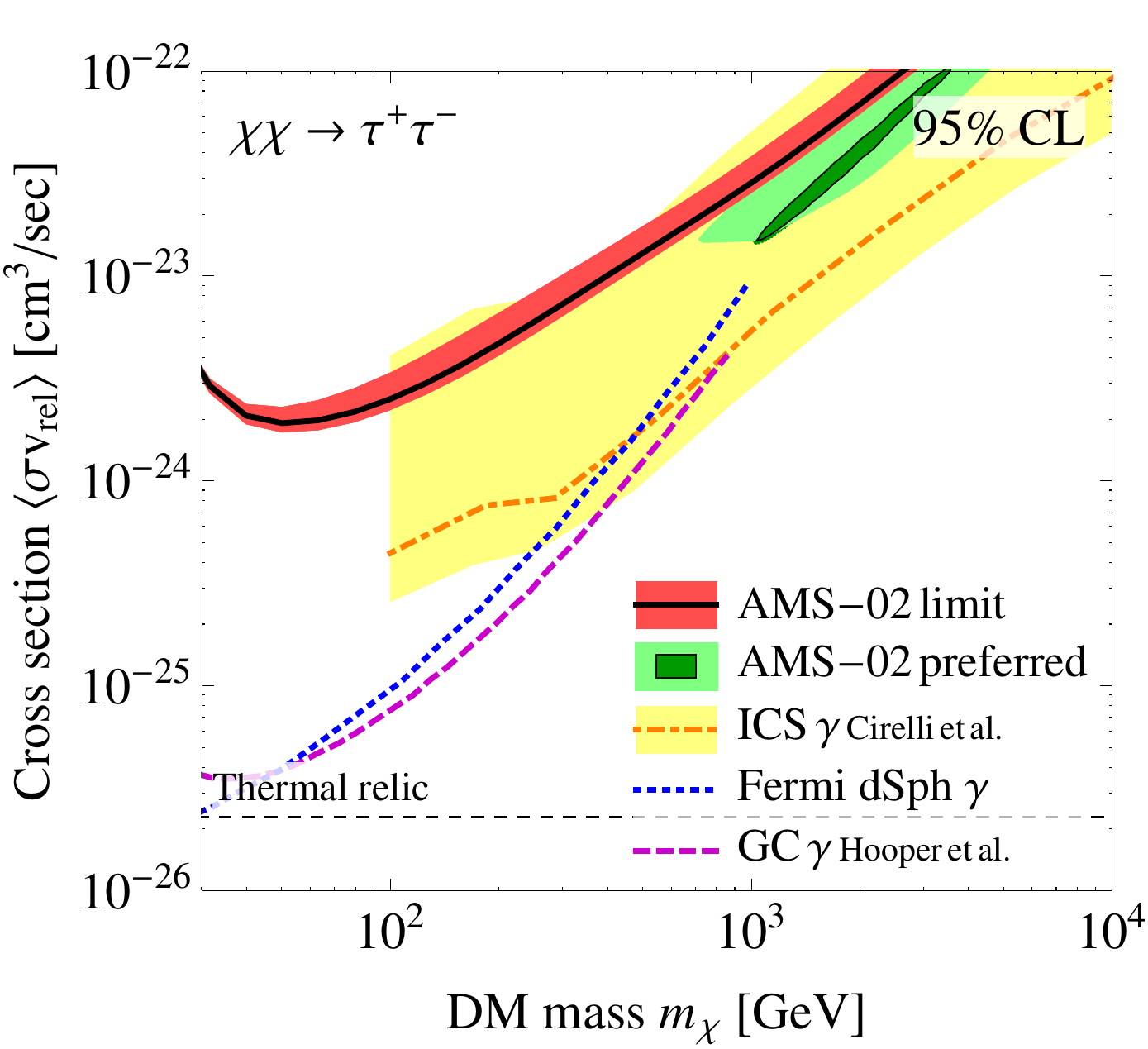} \\
    \multicolumn{2}{c}{\includegraphics[width=0.40\textwidth]{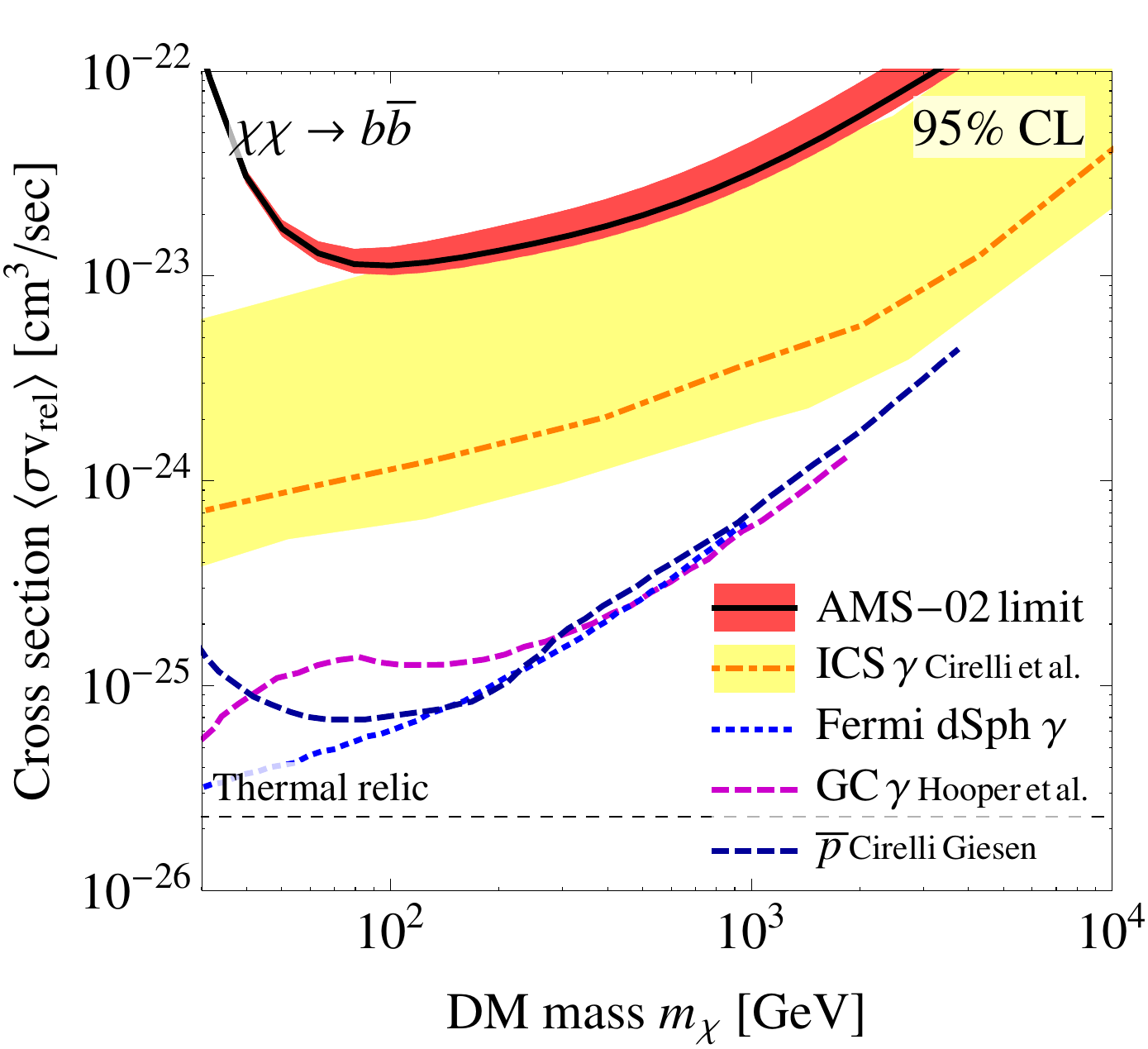}} \\
    \includegraphics[width=0.40\textwidth]{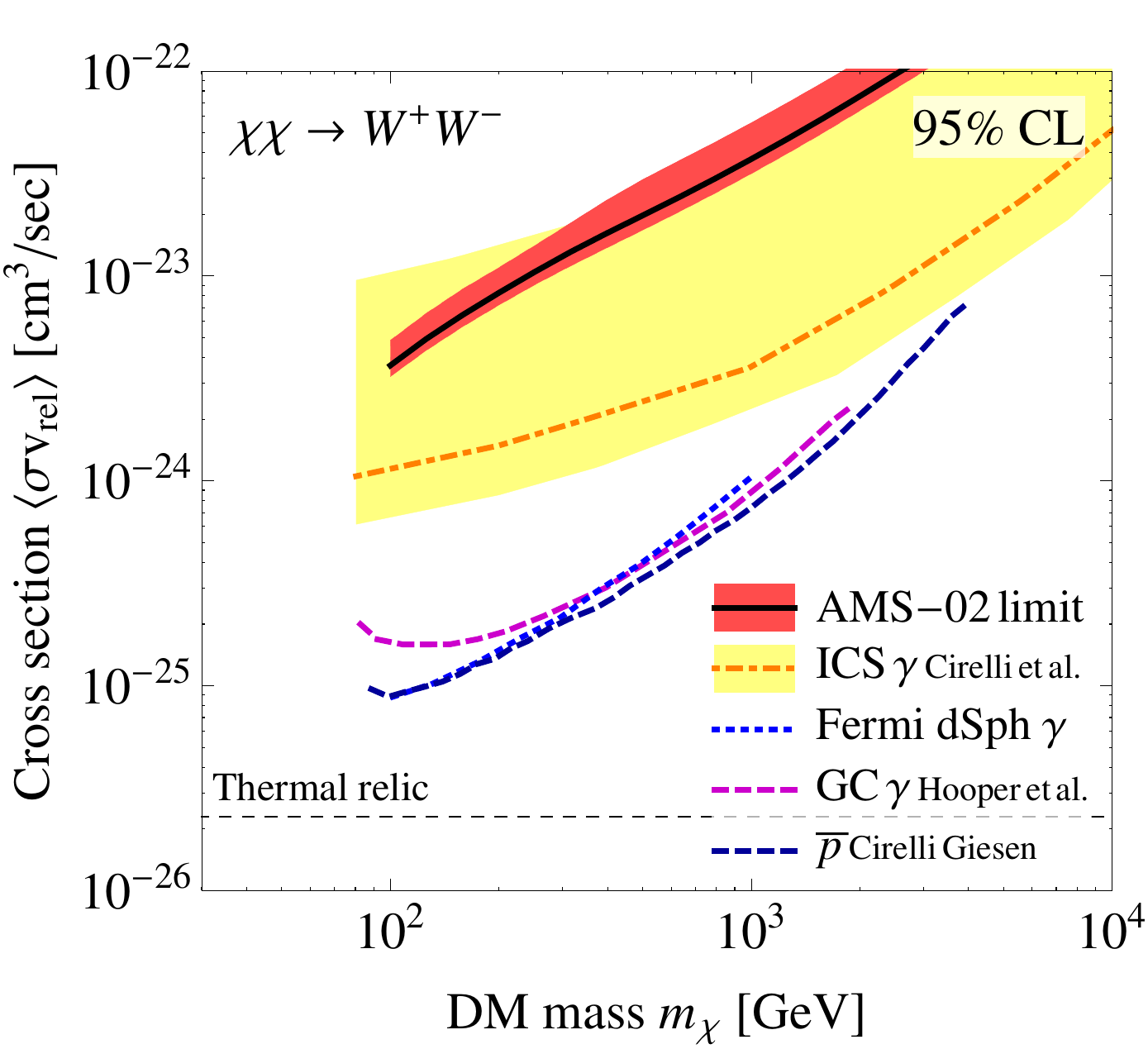} &
    \includegraphics[width=0.40\textwidth]{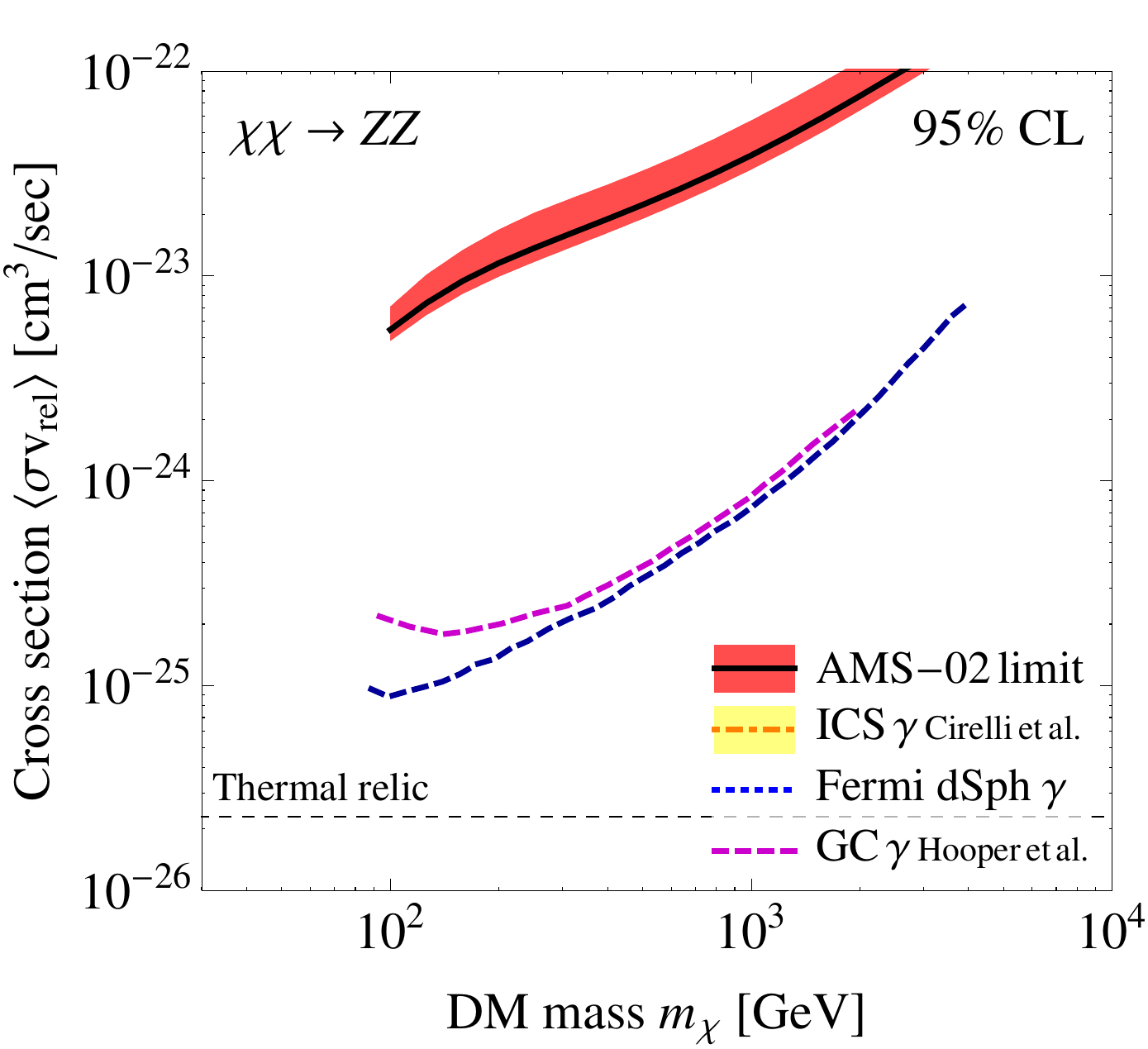}
  \end{tabular}
  \caption{Limits on the DM annihilation cross section $\ev{\sigma v_\text{rel}}$
    as a function of the DM mass $m_\chi$ for the
    annihilation channels
    $\chi\chi \to \mu^+\mu^-$ (top left),
    $\chi\chi \to \tau^+\tau^-$ (top right),
    $\chi\chi \to b\bar{b}$ (middle),
    $\chi\chi \to W^+W^-$ (bottom left),
    $\chi\chi \to ZZ$ (bottom right).
    Solid black lines show constraints derived in this work from AMS-02
    positron data, and red bands indicate how uncertainties in the
    positron propagation model~\cite{Cirelli:2010xx} affect
    these constraints. We have assumed an NFW profile for the DM distribution in
    the Milky Way, but have checked that alternative choices lead to
    almost identical limits.
    Where available, we show also for comparison limits from Fermi-LAT
    observations of $\gamma$ ray emission in dwarf galaxies~\cite{Ackermann:2011wa}
    (light blue dotted)
    and in the Galactic Center~\cite{Hooper:2012sr} (purple dashed),
    from an analysis of PAMELA antiproton data~\cite{PAMELA:2010rc,Cirelli:2013hv}
    (dark blue dashed),
    and from inverse Compton scattering~\cite{Cirelli:2009dv} (orange line = NFW
    profile, lower (upper) edge of yellow band = Einasto (isothermal) profile).
    The horizontal dashed line shows the annihilation cross section that yields the
    correct DM abundance via freeze-out~\cite{Steigman:2012nb}. For $\chi\chi \to \mu^+\mu^-$
    and $\chi\chi \to \tau^+\tau^-$
    we indicate in green the parameter regions that would be favored by attempts
    to explain the positron excess in terms of DM annihilation (dark green = background
    and propagation models based on Fermi observations, light green =
    MIN, MED, MAX propagation models).}
  \label{fig:limits}
\end{figure*}

\paragraph*{Results and discussion.} In figure~\ref{fig:limits} we show the limits on the DM
annihilation cross section $\ev{\sigma v_\text{rel}}$ and the DM mass $m_\chi$
derived from AMS-02 data. ($\ev{\sigma v_\text{rel}}$ denotes the annihilation
cross section, multiplied by the relative velocity of the two annihilating DM
particles and averaged over their velocity distribution.) Solid black lines
correspond to the MED propagation model, whereas red bands
indicate the difference between the MIN and MAX propagation models. We have
assumed the DM distribution $\rho(r)$ in the Milky Way to follow the
NFW profile~\cite{Navarro:1995iw} $\rho_\text{NFW} = \rho_s (r_s / r) (1 + r/r_s)^{-2}$
with the parameters $\rho_s = 0.184$~GeV/cm$^3$, $r_s = 24.42$~kpc~\cite{Cirelli:2010xx},
%Einasto profile~\cite{Graham:2005xx,Navarro:2008kc} $\rho_\text{Einasto} = \rho_s
%\exp\big(-2/\alpha \times [(r/r_s)^\alpha - 1] \big)$ with the parameters
%$\rho_s = 0.033$~GeV/cm$^3$, $r_s = 28.44$~kpc, $\alpha =
%0.17$~\cite{Cirelli:2010xx},
but we have verified that the choice of
halo profile has a negligible impact on our limits. The reason is that high
energy positrons cannot travel too far in the Milky Way before losing energy,
so that the flux observed at the Earth has to come from our local
galactic neighborhood, where the DM halo profile has relatively small
uncertainties. 
We compare the AMS-02 limits to the bounds obtained by
Cirelli and Giesen from the antiproton fluxes measured in
PAMELA~\cite{PAMELA:2010rc, Cirelli:2013hv}, to the bounds obtained by the
Fermi-LAT collaboration from an analysis of gamma ray signals from dwarf
galaxies~\cite{Ackermann:2011wa} (see also~\cite{GeringerSameth:2011iw}),
to the bounds obtained by Hooper et al.\ using Fermi-LAT gamma ray observations
of the Galactic Center~\cite{Hooper:2012sr} (see also~\cite{Crocker:2010gy,
Bertone:2008xr, Regis:2008ij}), and to limits from inverse Compton
scattering (ICS)~\cite{Cirelli:2009dv}, again based on Fermi-LAT data.
Note that the antiproton constraints are
based on a particular cosmic ray propagation model and can
vary by more than an order of magnitude if the propagation parameters are
changed. Gamma ray signals from the Galactic Center on the other hand depend very
strongly on the details of the DM distribution in that region. We use here
the most conservative limits from~\cite{Hooper:2012sr}, based
on the assumption of a cored profile. Significantly stronger limits are
obtained for steeper profiles, which are also in better agreement with
simulations of galaxy formation and evolution (see~\cite{Hooper:2012sr}
for details).  For the ICS limits, we depict the uncertainty in the DM density
distribution as yellow bands.  The limits are taken from~\cite{Cirelli:2009dv}
and are based on Fermi data from a $3^\circ \times 3^\circ$ region around the
Galactic Center.  ICS exclusion limits derived from larger regions of interest
are less dependent on the DM density distribution and would lie close to the upper
ends of the yellow bands in fig.~\ref{fig:limits}. Note that IceCube limits on
neutrinos from DM annihilation~\cite{Abbasi:2011eq} are still outside the
parameter range shown in fig.~\ref{fig:limits}, but can be expected to improve
with more statistics.

We see that for most annihilation channels, AMS-02 constraints are weaker than
bounds from PAMELA and Fermi-LAT. Only for annihilation to $\mu^+ \mu^-$,
AMS-02 and Fermi-LAT $\gamma$ ray limits can be comparable, but only when very conservative
assumptions on the DM halo profile are made to avoid ICS bounds.
The reason why hadronic annihilation channels are more strongly constrained by
Fermi-LAT data is the occurrence of prompt $\gamma$ rays from $\pi^0$ decay.
Antiproton constraints are not
competitive in the $\mu^+ \mu^-$ and $\tau^+ \tau^-$ channels and are therefore
not included in figure~\ref{fig:limits} for these channels.  Fermi-LAT results are not available
for the $Z Z$ final state, but are expected to be similar
to those for the $W^+ W^-$ final state~\cite{Bertone:2008xr}.

\begin{figure}
  \begin{center}
    \includegraphics[width=0.45\textwidth]{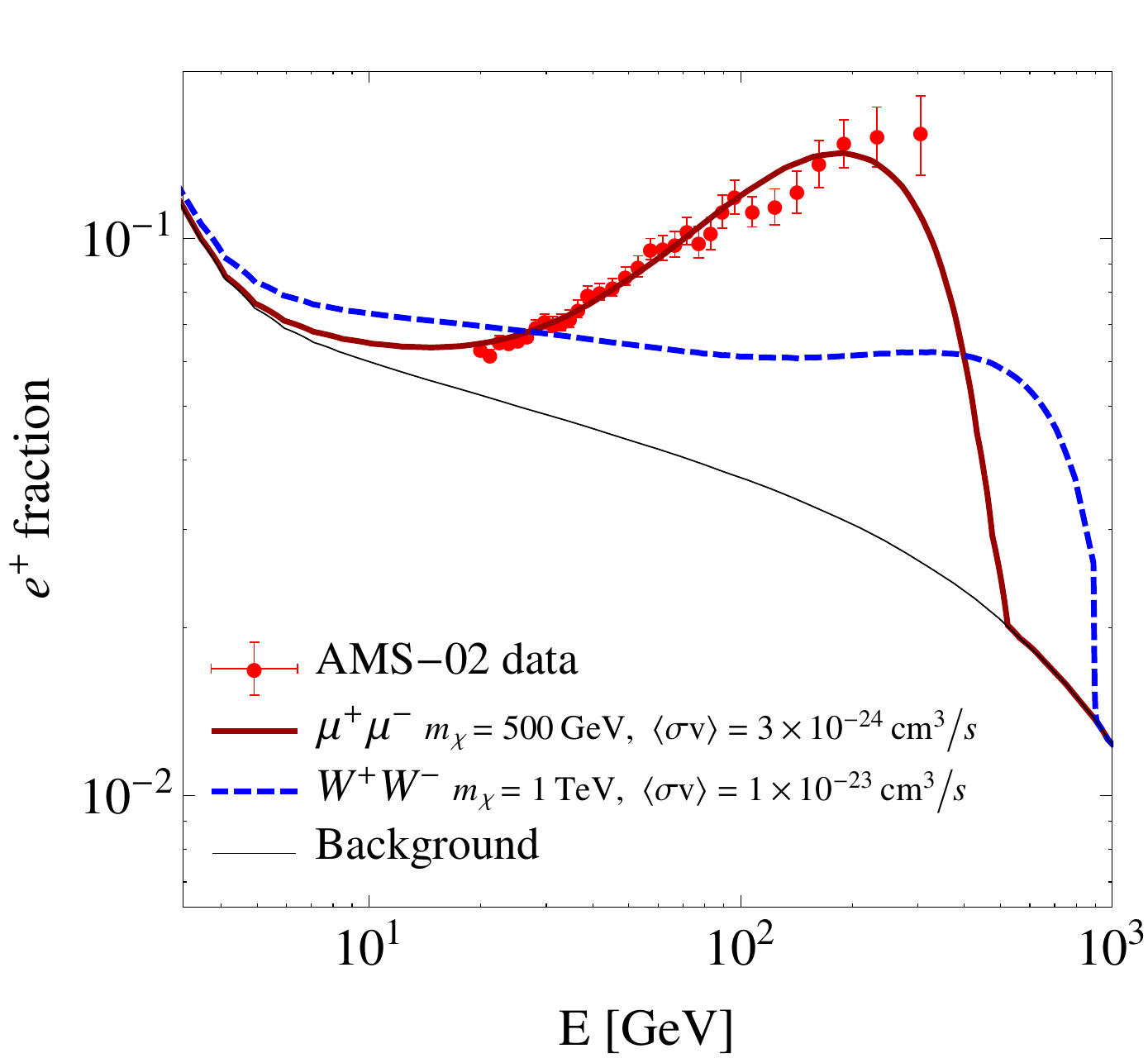}
  \end{center}
  \caption{Comparison of the positron fraction measured by AMS-02
    to two particular dark matter scenarios.}
  \label{fig:spectrum}
\end{figure}

For the $\mu^+ \mu^-$ and $\tau^+ \tau^-$ final states, we also illustrate in
figure~\ref{fig:limits} the values of $m_\chi$
and $\ev{\sigma v_\text{rel}}$ that would be needed to \emph{explain} the
positron excess in terms of dark matter (green shaded regions).
For the $b\bar{b}$, $W^+ W^-$ and $ZZ$ final states, we find that
the predicted positron spectrum is typically too flat to explain the data.
This is also illustrated in figure~\ref{fig:spectrum}, where we compare
several specific DM scenarios to AMS-02 data.
We conclude from figure~\ref{fig:limits} that interpreting
the observed $e^+$ excess in terms of dark matter remains difficult.
On the one hand, constraints from antiproton and gamma ray observations
are extremely constraining, especially for the $b\bar{b}$, $W^+ W^-$
and $ZZ$ final states. We emphasize again that the Galactic Center
gamma ray limits shown in figure~\ref{fig:limits} are very conservative,
and significantly stronger limits are obtained for less conservative
assumptions on the dark matter profile in the Galactic Center region.
Even if the constraints are avoided, the annihilation cross section required
to explain the positron data is significantly larger than the
thermal relic value $\ev{\sigma v_\text{rel}} \simeq 2.3 \times
10^{-26}$~cm$^3$/sec~\cite{Steigman:2012nb} (horizontal dashed line in
figure~\ref{fig:limits}). Mechanisms to enhance $\ev{\sigma
v_\text{rel}}$ today while retaining a smaller value in the early Universe
include Sommerfeld enhancement~\cite{Hisano:2006nn, ArkaniHamed:2008qn} and
resonant enhancement~\cite{Ibe:2008ye}.  In both of these scenarios,
$\ev{\sigma v_\text{rel}}$ acquires a velocity dependence, which leads to small
cross sections at the time of DM freeze-out ($\ev{v_\text{rel}^2} \simeq 0.24
c^2$), but larger cross sections in the Milky Way today ($v_\text{rel} \sim
\text{few} \times 100\ \text{km/sec}$).  In dwarf galaxies, DM
velocities are even smaller, of order $\text{few} \times 10\ \text{km/sec}$.
Fermi-LAT gamma ray limits from dwarf galaxies will impose severe
constraints on such models because of the small velocity dispersion in
dwarf galaxies. Constraints are typically weakest for final states composed
exclusively of light charged leptons: antiprotons are not produced in this
case, and the expected gamma ray flux is significantly weaker than for final
states containing hadrons. An interesting possibility in this context is DM
annihilating to a light intermediate state which then decays into
leptons~\cite{Mardon:2009rc}, with decays to heavier particles forbidden by the
low mass of the intermediate state.  Other possibilities are that DM is
non-thermally produced in the early Universe (for instance through decays of a
long-lived intermediate state), or that there are several dark matter
components: a dominant one that drives galactic dynamics, and a subdominant one
that exhibits stronger clustering (for instance due to relatively large
self-interactions), leading to locally larger densities (see for
instance~\cite{Fan:2013yva}).

\paragraph*{Conclusions.} We have derived limits on dark matter annihilation
from the cosmic ray positron fraction measured by AMS-02.  Due to the yet
unexplained positron excess, these limits are in most cases not competitive
with constraints from gamma ray and antiproton observations.  On the other hand,
models capable of \emph{explaining} the positron excess through
DM annihilation are severely constrained experimentally by the non-observation of anomalous
gamma ray and antiproton populations, and theoretically by the requirement of
extremely large annihilation cross sections. Fitting AMS-02 data in such models may be
marginally possible if the annihilation is exclusively to leptons, and if
very conservative assumptions on the DM profile in the Galactic Center
region are adopted to weaken $\gamma$ ray constraints.

{\emph{Note added:} After the first version of this manuscript appeared on the
arXiv, several studies have found possible tension between the AMS-02 positron
fraction and Fermi-LAT data on the $e^+ + e^-$ spectrum, see for
instance~\cite{Cholis:2013psa, Yuan:2013eja, Feng:2013vva}. In particular, it
is difficult to explain both data sets simultaneously with a new population of
high-energy electrons and positrons (coming e.g.\ from DM annihilation or decay
or from an astrophysical source), as long as this population is
charge-symmetric.  It is at the moment unclear whether the tension is due to a
problem with the propagation model for electrons and positrons, due to an
underestimated systematic effect in one of the experiments, or due to a charge
asymmetry in the new $e^+ e^-$ population~\cite{Masina:2011hu, Feng:2013vva}.
If future studies should reveal that the tension is due to experimental systematics,
also the uncertainties in our procedure of extracting the positron flux by combining
AMS-02 and Fermi data would increase.

\paragraph*{Acknowledgment} I would like to thank CERN for kind hospitality and
support during part of this work.

% Bibliography
\bibliographystyle{apsrev}
\bibliography{./ams}

\end{document}